\documentclass[twoside,12pt]{article} 
\begin{document} 
\pagestyle{myheadings} 
\markboth{AADEBUG 2000}{Execution replay and debugging} 
\title{Execution replay and debugging\footnote{In M. Ducass\'e (ed),
proceedings of the Fourth International Workshop on Automated
Debugging (AADEBUG 2000), August 2000, Munich. COmputer Research
Repository (http://www.acm.org/corr/), cs.SE/0011006; whole
proceedings: cs.SE/0010035.}}

\author{
Michiel Ronsse$^{1,}$, 
Koen De Bosschere$^{1,}$\\
and Jacques Chassin de Kergommeaux$^{2,}$\\
\small$^1$ELIS, Ghent University, St.-Pietersnieuwstraat 41, B-9000 Ghent, Belgium\\
\small$^2$ID-IMAG, B.P. 53, F-38041 Grenoble Cedex 9, France
}

\date{} 
            
\maketitle 
            
\begin{abstract} 
As most parallel and distributed programs are internally non-determi\-nistic --
consecutive runs with the same input might result in a different
program flow -- vanilla cyclic debugging techniques as such are
useless. In order to use cyclic debugging tools, we need a tool that
records information about an execution so that it can be replayed for
debugging. Because recording information interferes with the execution, we
must limit the amount of information and keep the processing of the information
fast. This paper contains a survey of existing execution replay
techniques and tools.
\end{abstract}

\section{Introduction} 

Conceptually, a parallel or distributed program\footnote{For the
remainder of this paper, we will use the term `{\em parallel} program'
for a program consisting of a number of processes running on a
multiprocessor with shared memory and the term `{\em distributed}
program' for a program running on a number of computers without shared
memory.} consists of a set of cooperating processes that are executed
in parallel by different processors. Writing parallel programs is
generally considered more difficult and more error-prone than
writing sequential programs as one has not only to concentrate on the
implementation of a particular algorithm, but also on the communication
and synchronisation between the processes.  Most contemporary parallel
programming tools are focused on automatic parallelisation of sequential
programs, or the analysis and visualisation of parallel programs, not on
the development of hand-written parallel or distributed programs. 
Consequently, there
is a clear lack of development tools for parallel and distributed programs.

A standard debugger that is used for `cyclic debugging', i.e.\ re-executing
the program over and over again with the same input, and zooming in on the  program 
execution until the bug is found (using break points, watch points, etc.) assumes
that a program can be re-executed deterministically.
This is clearly not the case for many parallel and distributed programs.
Indeed, a standard source level debugger will change the timing of the processes,
and hence maybe alter the program flow. As a result, it can cause the symptoms of a particular
bug to suddenly disappear, or to be replaced by other symptoms
(so-called Heisenbugs~\cite{Ledoux1}).

In order to be able to use the wealth of debugging tools that have been
developed for sequential programs for the debugging of parallel programs,
we need a way to deterministically re-execute a parallel program.
The main problem is that parallel programs are non-deterministic (especially the faulty ones):
each program run (even with the same input) might result in a different
program execution. Although non-determinism is also present in
sequential programs (caused by interrupts, etc.) its presence is far
more visible in parallel programs because a lot of non-deterministic
constructs are used on purpose: e.g., the order in which processors
use a semaphore is not planned by the program code but is determined by the
competition between the different processes.

One way to enable cyclic debugging techniques for parallel or
distributed programs is by the usage of the so-called {\em execution
replay} technique. This technique conists of two phases: first a
trace of a parallel execution is made (record phase), and afterwards
the trace is used to control the re-execution of the program (replay
phase)\footnote{A comparable technique is on-the-fly replay, as
described in~\cite{gerstel}. This system uses two parallel computers:
an execution is recorded on the first one and is replayed {\em at the
same time} on the second computer.}, provided one can supply the same
input to the program as the one supplied during the recorded run: both
interactive input (keyboard) and file input should be identical and
also system calls should return the same result (e.g.\ time-of-day,
system usage, \ldots).  Since the re-execution is now deterministic,
intrusive cyclic debugging tools can be used to debug the program:
visualisation, data race detection, etc.\ is possible during replay
without perturbing the original program execution.

In this survey, we start with an overview of the main sources of
non-determi\-nism in sequential, parallel and distributed programs,
followed by a section on the main issues in execution replay
a section on logical clocks,
and two sections describing the execution replay methods that are
described in literature. Finally, the paper is concluded with a
conclusion, and an extensive list of references on this topic.

\section{Non-determinism} 
In order to be able to deterministically re-execute a program, we first
need to determine the non-determinism in a program execution. We start
by making a distinction between different sources of non-determinism.
\begin{itemize}
\item {\em External non-determinism} means that an application returns different
results for repeated executions with the same
input~\cite{Emrath1}. This kind of non-determinism can be desired or
undesired.  In the cases were it is not desired, it has to be
considered a bug that has to be removed from the program. In the other
cases, e.g., a program that returns one solution for the eight queens
problem (which has different possible solutions)\footnote{Eight queens
are to be placed on a chess board in such a way that no queen
threatens any other queen.} this non-determinism has to be considered
a feature of the program instead of a bug.
\item {\em Internal non-determinism} means that repeated executions with the same
input yield the same result, but the internal execution path is
different. This allows to exploit different alternative executions,
balancing the load, maximising the potential parallelism in the implementation, etc.
\end{itemize}                                                         

The amount of internal non-determinism depends on the level of abstraction.
E.g.\ a program can be internally deterministic at the level of 
semaphore operations, but not at the level of e.g., mutexes or spinning
loops used to implement the semaphores. It turns out that the amount
of internal non-determinism increases with lower abstraction levels. If 
a program is internally deterministic at the highest abstraction level,
it will --by definition-- be externally deterministic too.

In order to be able to re-execute an execution deterministically, one
should know all the possible causes of non-deterministic behaviour. The next
sections describe the different causes for sequential, parallel
and distributed programs.

\subsection{Sequential programs} 
A sequential program execution is determined by the program code\footnote{This
paper assumes that the code is not changed during the execution, hence
self-modifying code is not dealt with.} and the program input. The
input can be provided at program start (initial values of the memory
locations) or during the execution (input from disk, keyboard,
network, etc.).

Re-execution can only be deterministic if the input to the program is exactly the
same as during the original execution. Replaying the input from keyboard,
disk, network, and so on, is fairly straightforward to implement by
logging the data during the first execution and refeeding it during
the re-executions. This also applies to some system calls such
as {\tt gettimeofday()}.

Some sources of input are harder to replay because not only the data
should be refed, but this should be done at the correct moment. This
applies to interrupts or signals because they cause a program
transfer. This input can only be faithfully replayed by logging the
moment of the interruption as well, e.g.\ using some kind of SIC
(Software Instruction Counter) as in~\cite{KA94-12,schuster}.

This paper will not deal with the non-determinism caused by these types
of input but will only focus on the additional non-determinism caused
by the parallel or distributed nature of an application.

\subsection{Parallel programs} 

For parallel programs one should also take into account the `internal'
input operations. Indeed, one could consider the memory read
operations as input operations as they potentially read a value
written by another process. If one executes a parallel program on a
monoprocessor, we can even consider the scheduler as the
cause of non-determinism instead of the many read operations. Indeed,
for these machines the scheduler operations determine the program
execution: another execution with the same scheduling will lead to the
same execution. However, this requires a scheduler that intervenes at
exactly the same moment as during the first execution.

As the non-determinism caused by memory operations is an important
concept for parallel programs, this type of non-determinism is called a
`race condition'. More specifically, a {\em race condition} is defined
as two unsynchronised accesses to the same shared location, and at
least one access modifies it. We have to make a distinction between
two types of race conditions: race conditions that are used to make a
program intentionally non-deterministic: {\em synchronisation races},
and race conditions that were not intended by the programmer ({\em
data races}).

We need {\em synchronisation races} to allow for competition between
threads to enter a critical section, to lock a semaphore, or to
implement load balancing.  Removing synchronisation races makes a
program completely deterministic. Therefore, in this paper, we do not
consider synchronisation races a programming error, but a functional
and useful characteristic of a parallel program.

{\em Data races} are not intended by the programmer, and are mostly
the result of improper synchronisation. By changing the
synchronisation, data races can always be removed.  It is
important to notice that the distinction between a data race and a
synchronisation race is actually a pure matter of abstraction. At the
implementation level of the synchronisation operations, a
synchronisation race is caused by a genuine data race (e.g., spin
locks, polling, etc.) on a synchronisation variable.

\subsection{Distributed programs}

For distributed programs, non-determinism is mainly introduced by
so-called {\em promiscuous receive operations} (e.g., {\tt
MPI\_Recv(...,MPI\_ANY\_SOURCE,...)} for MPI~\cite{MPI} programs) which can
receive a message from any other process. As the source of the message
is not specified, it is possible that another message is received
during a re-execution. In a sense, they play the role of
`racing' store operations in non-deterministic programs on
multiprocessors.

We assume that messages in a point-to-point communication comply with
the non-overtaking property which means that successive messages sent
between two nodes are always received in the order in which they were
sent (this is the case for MPI and PVM~\cite{Geist}). Hence, since messages are
assumed to be produced deterministically, receive operations in a
private point-to-point communication that specify the sender are always
deterministic.

Besides the promiscuous receive operations, there is another class of
instructions that can cause non-determinism in a message passing
program: test operations for non-blocking message operations. These
non-blocking operations return to the caller immediately: they do not
wait until the message was received/delivered. In this case, test
operations are used to check for the arrival of messages or to check
if a send operation has finished. Non-blocking test operations are
intensively used in message passing programs that use PVM or MPI,
e.g., to maximally overlap communication with computation.

By the very fact that the test operations are non-blocking, they can
be used in polling loops. The actual number of calls will depend on
timing variations of parallel program, and is thus
non-deterministic. Although many programs will not base their
operation on the number of failed tests, some could do so (e.g., to
implement a kind of time-out), and hence cannot be correctly replayed
when not recorded.

\section{Main issues in execution replay} 

For record/replay systems one will record the non-determinism at a 
particular abstraction level, and enforce these non-deterministic choices at the
same abstraction level during the re-execution. This means that all events that happen on a 
higher abstraction level will be replayed faithfully, while nothing is 
known about the events on lower abstraction levels.

On the other hand, 
the level of abstraction will also determine the amount of information 
needed to allow for a faithful replay. The lower the abstraction level,
the more information about the original execution we can obtain for 
the debugging session, but the more time and space we need to record
the non-deterministic choices (hence more perturbation). The higher the
abstraction level, the lesser we have to record, but the more uncertainty
we have during debugging about the original execution. Determining the 
level of abstraction is hence of paramount importance and can make a 
record/replay system either practical or impractical. In practice,
one is only interested in a re-execution that is
equivalent `as far as the programmer is concerned', meaning that he/she
wants to have access to his/her own code, but he/she is not interested
in the code of the libraries he/she is using, as long a the 
semantics of the library calls is preserved. An equivalent re-execution
of the implementation of library routines is not needed, as the programmer
cannot observe -let alone debug- these routines anyway.

A practical execution replay system should satisfy two properties, as
described in the next two sections.

\subsection{Low overhead of the record phase}

The recording overhead must be low in
time~\cite{ATHA-01} and in space~\cite{Leblanc,Leu2,Netzer93}.

The {\em time overhead} should be low to circumvent Heisenbugs and to
limit the probe effect~\cite{Gait2}. A low overhead ensures that the
recorded execution is more or less equivalent with a regular execution
(without tracing). If the time overhead is low enough we can leave the
tracing turned on all the time, even in production code. The trace can
then be a standard part of a bug report.  Even with a zero overhead, a
program execution in record mode, is not guaranteed to be identical to
the previous execution without trace mode, because no two executions
are guaranteed to be identical in the presence of internal
non-determinism. Leaving the tracing on all the time is the only way
to guarantee a correct replay of a faulty execution.

The {\em space overhead} for the trace files should be limited
too. The first reason is that this is necessary to be able to trace
long running programs. The second reason is that storing the trace
requires some bandwidth, which should be shared with the target
program. So, storing more means perturbating more. Therefore, the
lesser is recorded, the better it is, as long as it still allows a
faithful replay. As the elimination of redundant information must be
done at run-time (and in real-time) the analysis should be as
simple as possible, minimising program perturbation. It is clear that
there is a tradeoff between the space and the time overhead. The space
overhead can be limited by using the already mentioned on-the-fly
replay method or by using {\em incremental replay}
techniques~\cite{Netzer94}. Incremental techniques use a combination of
checkpoints with execution traces. These execution traces only contain
information about the execution since the last checkpoint was
taken. Replay is then only possible for the part of the execution
after this checkpoint, making it hard to find bugs for which there is
a long time between the occurrence of the bug and the time at which the
bug starts to have an effect on the execution. Furthermore, taking
consistent (distributed) checkpoints is not that easy.

\subsection{Faithful re-execution}
The necessary condition for using cyclic debugging is that we
can re-execute a program as many times as needed, and that the 
re-executions are in some way `equivalent' with the original
execution. As explained above, the abstraction level will be
a critical issue.  Therefore, an important question for
an execution replay method is what and how much to record during the
recorded execution.

There are two possible approaches to force executions of a parallel or
distributed program to be equivalent to a traced execution. The first
one is to force the processes to read the same values of shared
variables or to receive the same messages as during the traced
execution by recording the original value ({\em content-based or
data-based replay}). The second one ({\em ordering-based replay})
makes sure that the interactions between the different processes occur
in the same order as during the original execution. For parallel
programs, the processes are forced to access the shared variables in
the same order as during the traced execution, forcing the variables
to undergo the modifications in the same order as during the record
phase. For distributed programs, the processes are forced to receive
the messages in the same order as during the original execution.

In this case, a scheme for detecting and recording the order of the
operations should be used. For detecting the order in which operations
are executed (or for detecting that there is no ordering and the
operations are therefore in parallel), logical clocks are commonly
used.

\section{Logical clocks}

A logical clock~\cite{raynal92,raynal96} $C()$ should obey the
so-called clock condition\footnote{$a \rightarrow b$ means that
operation $a$ `happened before' operation $b$ meaning there is some
sort of causal relation between the two events.}
\[a \rightarrow b \Rightarrow C(a) < C(b) \] 
This relation simple states that if $a$ occurs {\em causally} before
$b$, the timestamp of $a$ should be smaller than the timestamp of
$b$. As it is sufficient to order the subsequent operations on the
{\em same} variable, such a clock normally calculates a new timestamp
based on the old timestamp of the process and the timestamp attached
to the last operation on the variable. It is clear that it is not
sufficient to use the wall time of the operations as timestamps as the
wall clock on different processes (especially in distributed machines)
is not synchronised.

The simplest form of a logical clock is the scalar Lamport
clock~\cite{LamClock}: a scalar value is attached to each
process. Each time a process executes an operation, the clock produces
a new value: the new clock is the successor of the maximum value of
the last timestamp of the process {\em and} of the timestamp attached
to the last operation on the same object. This is a fairly natural way
of updating the clock: the new operation happens after the last
operation on the object and the last operation of the process; hence
the new timestamp should be bigger than the timestamps attached to
these two operations. It is clear that this type of clock satisfies
the clock condition. If we have two operations $a$ and $b$ with
timestamps $C(a)$ and $C(b)$ then we have three possibilities:
\begin{itemize}
\item $C(a)=C(b)$ meaning that there is no causal relation between 
these two operations.
\item $C(a)<C(b)$ meaning that there could be a causal relation between
these  operations ($a \rightarrow b$), but we are sure that the
contrary relation ($b \rightarrow a$) is not true.
\item $C(a)>C(b)$ meaning the same as above with $a$ and $b$ switched.
\end{itemize}
Note that this type of clock provides enough information to get a
faithful replay: if we trace the Lamport timestamps of the
operations, we get a correct replay if we stall an operation $x$ with
timestamp $C(x)$ until all operations $y$ with $C(y)<C(x)$ have been
executed.

Vector clocks~\cite{fidge,Mattern} are used if one wants to obtain
more information about an execution. A vector clock for a program with
$N$ processes consists of $N$ scalar values. If process $p$ executes
an operation, the new vector timestamp is calculated as follows:
first, a new vector timestamp is calculated as the supremum of
the last timestamps of the process and the object on which the
operation is performed, and then the $p$-th element is incremented.  An
interesting property of a vector clock is that it not only satisfies
the clock condition but also the stronger condition (meaning they are
strongly consistent) \[a \rightarrow b
\Leftrightarrow C(a) < C(b)\] This means that given two vector clocks
$C(a)$ and $C(b)$, it is possible to deduce the causal relation
between the two operations $a$ and $b$.

It is possible to augment the dimension of the logical clock even
more, resulting in matrix clocks~\cite{wuu, sarin}.  Matrix clocks 
provide second order information to a process. It is a list of 
vector clocks, namely per process the last vector clock that was
communicated to the current process.  This information can be used
to discard obsolete information in distributed systems.

\section{Execution replay methods for parallel programs} 

In this section, an overview of the most important replay methods
described in literature is given.

\subsection{Content-based}
In~\cite{VAX} a content-based replay method (name {\em Recap}) that
traces the data read from every shared memory location is
proposed. A trace generation of 1 MB/s on a VAX 11/780 was measured, 
making content-based replay impractical as the time needed to record
the large amount of required information is significant, which might
modify the initial execution considerably. Moreover, not only the time 
but also the space overhead is too large.

This replay method has however one benefit: it is possible to replay a
subset of the processes (or even one process) in
isolation. Nevertheless, it is argued~\cite{Leblanc} that this is no
real benefit as it is difficult to examine the interactions between
the different processes, hindering the task of finding the cause of a
bug. Today, content-based replay is only used for tracing I/O and for
tracing the result of certain system calls such as {\tt
gettimeofday()}. Indeed, this is the only viable solutions as it is
not possible to `replay' the operations that produced the result of
these operations.

\subsection{Ordering-based}
As mentioned above these methods try to guarantee that the shared
memory dependencies during the replay phase match the dependencies that
were observed during the record phase. This will make sure that e.g.\
each {\em read} is executed after the same {\em write} as during the
recorded execution forcing the same data to be read. It turns out that
this approach allows for a dramatical reduction in the time and space overhead
in the recording phase.
The tracing of the order of the memory operations can be
done using hardware~\cite{Bacon1}, software or hybrid probes.

\subsubsection{Replaying on a monoprocessor}
Replay mechanisms based on the scheduling order of the different
threads can be used for monoprocessor systems. Indeed, by imposing the
same scheduling order during replay, an equivalent execution is
constructed~\cite{holloman, russ, schuster}. This scheme can be
extended to multiprocessor systems by also tracing the memory
operations executed between two successive scheduling
operations. In~\cite{choi-java}, such an implementation for
Java is described. Since a typical execution of a Java program has a small number
of scheduler operations (no time slicing is used and therefore
scheduling is only performed at predefined points such as {\tt
monitorenter} calls) they succeed in producing very small trace files
albeit at the cost of a large overhead (17-88\%).

Such a method is of course only viable if one has complete control
over the operations of the scheduler, as is only the case for
proprietary systems or virtual machines such as the JVM used by Java
applications.

\subsubsection{Replaying on a multiprocessor}
For systems where one has no control over the scheduler operations or
if one wants to replay on a multiprocessor more powerful replay
methods are needed. The methods described below are targeted at
non-proprietary systems such as UNIX systems and require no
modifications to the kernel: they all work in user space.

\paragraph{Instant Replay}
Instant Replay~\cite{Leblanc} is probably the most known replay
method. The method is targeted at CREW (concurrent reader, exclusive
writer) systems and attaches a {\em version number} to each shared
object. Each time a read operation is performed on an certain object,
the version number of the object is traced.  For each write operation
performed on an object, the version number of the object is incremented and
the number of read operations between the last two write operations is
traced. During the replay phase, each read operation stalls till the
version number of the object is correct, and each write operation
stalls until the same number of read operations for a specific version
(number) of the object are performed. It is clear that this method
only works if the memory operations obey the CREW property: Instant
Replay was proposed for a Butterfly with monitors.  If a process reads
or writes the object without using the monitor, the operation will not
be traced and a correct replay is impossible. As such, only the
synchronisation races are replayed, not the data races.

In~\cite{minas}, the Instant Replay method was adapted for a {\em
pSather}, a parallel object-oriented programming language.

\paragraph{Bacon \& Goldstein}
In~\cite{Bacon1}, a hardware-assisted replay method was
proposed. The system observers the cache traffic between the memory
and the CPU's and logs a subset of it. It is clear that this approach
is highly computer dependent but it introduces no overhead at all, at
the cost of extra hardware. Both the record and the replay phase use
additional hardware in order to accomplish this task.  A trace
bandwidth of 1.17MB/s was measured for a fine-grained shared memory
application on a 12-processor computer. 
As the tracing occurs at the lowest possible level, it is impossible
to distinguish the synchronisation operations from the `normal' memory 
operations. Therefore, this method replays both the synchronisation
races and the data races.

\paragraph{Netzer}
In~\cite{Netzer93-2, Netzer93} a replay system based on vector clocks
was presented. The system attaches a vector timestamp to each process
and each shared variable. Each time a process accesses a variable,
both the vector time\-stamp of the process and the object are
updated. As vector clocks are strongly consistent it is possible to
detect parallel operations on the same variable by comparing the
attached vector timestamps. If they turn out to be non-ordered,
the operations are not properly synchronised, and as
such the pair is involved in a race. By replaying the race in the same
order during the replay, the race causes no harm and the replayed
execution is equivalent with the recorded execution. The method
succeeds in limiting the space overhead because the vector clocks
automatically remove transitive ordered accesses from the trace.

Unfortunately, the method has a number of disadvantages:
\begin{itemize}
\item the size of a vector clock varies with the number of processes,
making it difficult to deal with programs that dynamically create
threads.
\item a vector clock has to be attached to all shared memory
locations. Moreover, the vector clocks have to compared at all shared
accesses. Although the papers about the method do not contain figures
about the time overhead, it is reasonable to expect a large overhead.
\item data and synchronisation races are treated equally: they are
both replayed. Although this sounds reasonable if one wants a correct
replay, we feel that a programmer would rather remove the data race
and replay the synchronisation races. After all, data races are often bugs
that should be removed from the program.
\end{itemize}

\paragraph{ROLT}
In~\cite{LL94-A, LL94-01} a method based on scalar Lamport timestamps
was proposed. Scalar timestamps are attached to each process and each
synchronisation operation. At each synchronisation operation new
timestamps are calculated. For a correct replay, it is sufficient to
stall each synchronisation operation until all synchronisation
operations with smaller timestamps have been executed. It turns out
that by logging the increments of the timestamps only, a substantial
reduction of the trace files can be obtained. It is clear that by
attaching scalar timestamps to all synchronisation operations, a total
order is imposed on these operations although the operations in the
recorded execution are only partially ordered. The additional
artificial dependencies are not harmful as they do not contradict the
actual execution order. They have however an impact on the replay
phase as it is possible that a synchronisation operation on one
synchronisation variable will have to wait for an operation on another
synchronisation variable.

By using so-called {\em fully snooped} variants~\cite{mr-ispan} of the
Lamport clocks it is possible to calculate a total order that is
consistent with the wall time of the operations, removing the
synthetic dependencies from the trace (at the cost of a bigger trace
file).

Compared to Netzer's method the system has the advantage that it
scales well (no vector clocks are used) and leads to small trace
files. Similar to Netzer's technique, the scalar clocks automatically
calculate the transitive reduction of the order of the synchronisation
operations.

The method was originally proposed for monitor operations (it was also
implemented for TreadMarks~\cite{TMK-00}, a distributed shared memory
system~\cite{MR-09}), making it unsuitable for programs containing
data races. In~\cite{MR-tocs} the replay method was extended with
automatic data race detection during the replay phase. 
During the record phase the order of the
synchronisation operations is recorded and during the first replayed
execution a data race detector runs as a watchdog. If a data race is
found, the replayed execution cannot be guaranteed to be correct after
the data race occurred, but this is no problem as the first data race
should be removed anyway. The data race can then be removed by using
cyclic debugging of the first part of the program. As soon as one is
sure that an execution no longer contains data races, a regular replayed
execution (without the data race detector running as a watchdog) can
be used to find the non-synchronisation related errors in the particular
execution. 

During the data race detection phase, information about all memory
operations is collected. As this would lead to a huge memory
consumption, snooped matrix clocks are used in order to remove
information that can be discarded (these are the memory operations
that can no longer race with new operations as all new operations are
causally related to the discarded ones.). The detection of the data
races themselves is performed using vector clocks.

\section{Methods for distributed programs} 
In the past, a whole deal of methods have been proposed for replaying
distributed programs. Implementing such a method for distributed
programs is simpler than for parallel programs as distributed programs
exchange information using message operations, and these operations
are typically part of a library or use a daemon (e.g.\ PVM and
MPI). Intercepting the operations is therefore straightforward as it
can be done by instrumenting function calls. This is not the case for
replaying parallel programs as individual memory operations have to be
traced.

Due to the large granularity one encounters in distributed
applications, the time overhead is not that important. Most methods
therefore primarily focus on reducing the space overhead.

Basically, two techniques are used. The first technique uses vector
clocks and was proposed by Netzer in~\cite{Netzer92}. The technique
basically checks whether  the send operation that corresponds to a receive
operation is ordered with the last receive operation of the same
process. If this is true, the messages cannot be received in the wrong
order, hence no race exists. If the receive and the send operations are
not ordered, this is traced. As was the case for Netzer's method for
parallel programs, this method also removes transitive orders from the
trace file. In~\cite{cc95} this method was implemented for
MPI-programs. The trace was augmented with information about the
number of non-blocking test operations.
Neyman et al. adapt this method for PVM-programs in~\cite{neyman}.
Unfortunately, 
non-blocking test operations are not dealt with, making an exact
replayed execution not always possible: any program that depends on
the number of test-operations performed (e.g., for time-outs) cannot
be correctly replayed.

The second technique is based on the fact that it is sufficient to
trace the actual sender of messages received by promiscuous receive
operations. This information can then be used during replay to force
the promiscuous receive operations to wait on a messages originating
from a particular sender process. Hence, a promiscuous receive
operation can be made deterministic during replay by replacing it
on-the-fly by a point-to-point receive operation. In~\cite{KranNOPE, mpl}
are some examples of this method. The first method does not deal with
non-blocking test operations leading to the same problems as mentioned
above. 

In~\cite{KrGr94} a completely different approach is taken. Instead of
replaying a recorded execution on an actual computer, the replay is
simulated. By altering the `transit time' of messages,
non-deterministic behaviour is tested. Another approach is used
in~\cite{andersson} where only one process is partially replayed
starting from a checkpoint. This method therefore offers a very
limited view on an execution. The system was implemented for PVM and
uses tagged messages in order to replay them.

In~\cite{MR-parco99}, a method for Athapascan~\cite{carissimi},
a hybrid parallel/distributed system is described. Athapascan consists of a number of nodes running on
different computers that communicate using messages (on top of
MPI). Each node consists of a number of POSIX threads communicating
using shared memory. 
The proposed replay method consists two parts. The previously
mentioned ROLT method is used for dealing with the non-determinism due
to the shared variables. For the promiscuous receive operations, the
actual sender of the messages received is recorded while the number of
test operations is logged for the non-blocking test operations.

\section{Conclusions}
It is clear that in the last 15 year, a lot of research and work has
been devoted to execution replay methods on behalf of the debugging
community. However, in order to get a perfect execution replay method,
one part is still missing: replaying the input. Although implementing
a replay tool for input seems straightforward --intercepting library
or system calls should suffice-- no tool exists at this moment,
probably because this is not exactly a research field requiring much
theoretical foundations.

\section*{Acknowledgements}
Michiel Ronsse is sponsored by a GOA project (12050895) from Ghent
University. Koen De Bosschere is a research associate with the Fund
for Scientific Research -- Flanders.

\end{document}